\documentclass[12pt]{article}
\usepackage[latin1]{inputenc}
\usepackage{amsmath, color}
\usepackage{amsfonts}
\usepackage{amssymb}
\usepackage[left=2.00cm, right=2.00cm, top=2.00cm, bottom=2.00cm]{geometry}
\usepackage{makeidx}
\usepackage{graphicx}
\usepackage{hyperref}
\usepackage{amsthm}
\usepackage{multirow}
\usepackage{multicol}

\usepackage[light,condensed,math]{kurier}
\usepackage[T1]{fontenc}

\usepackage{mathpazo}      % Math
\usepackage{charter}        % Text Charter
\newcommand{\FF}{{\mathbb{F}}}

\makeatother
\theoremstyle{definition}
\newtheorem{definisi}{Definition}[section]
\newtheorem{contoh}{Example}[section]
\newtheorem{catatan}{Remark}[section]

\theoremstyle{theorem}
\newtheorem{teorema}{Theorem}[section]
\newtheorem{lemma}[teorema]{Lemma}
\newtheorem{akibat}[teorema]{Corollary}
\newtheorem{proposisi}[teorema]{Proposition}
\newtheorem{alg}[teorema]{Algorithm}

\DeclareMathOperator{\ndo}{End}
\DeclareMathOperator{\sepan}{span}
\DeclareMathOperator{\homo}{Hom}

\DeclareMathOperator{\fpb}{gcd}

\newcommand{\pengN}{[\gamma A]{\fq^n}}
\newcommand{\pengM}{[A]{\fq^n}}
\newcommand{\fq}{\mathbb{F}_q}
\newcommand{\fsepergamma}{f^{1/\gamma}}
\newcommand{\fgamma}{f^\gamma}
\newcommand{\pgamma}{p^\gamma}
\newcommand{\bukti}{\textbf{Proof. }}

\newcommand{\fxmod}{$\fq[x]$-module }
\newcommand{\ndofx}{\ndo_{F[x]}}
\newcommand{\Fnn}{\fq^{n\times n}}
\newcommand{\homofx}{\homo_{F[x]}}
\newcommand{\homofqx}{\homo_{\fq[x]}}

\newcommand*{\QED}{\hfill\ensuremath{\blacksquare}}

\title{On bases and the dimensions of twisted centralizer codes}
\author{Ahmad Muchlis \footnote{Algebra Research Group, Faculty of Mathematics and Natural Sciences,
Institut Teknologi Bandung,
Jl. Ganesha 10, Bandung, 40132,
INDONESIA,},
Galih Pradananta \footnote{Algebra Research Group, Faculty of Mathematics and Natural Sciences,
Institut Teknologi Bandung,
Jl. Ganesha 10, Bandung, 40132,
INDONESIA,\newline
Current address: Department of Mathematics Education, Institut Agama Islam Negeri Tulungagung,
Jl. Mayor Sujadi Timur 46, Tulungagung, 66221,
INDONESIA},
Pudji Astuti\footnote{Algebra Research Group, Faculty of Mathematics and Natural Sciences,
Institut Teknologi Bandung,
Jl. Ganesha 10, Bandung, 40132,
INDONESIA,},
\hspace{0.1cm}and Djoko Suprijanto\footnote{Combinatorial Mathematics Research Group,
Faculty of Mathematics and Natural Sciences,
Institut Teknologi Bandung,
Jl. Ganesha 10, Bandung, 40132,
INDONESIA, \hfill{djoko@math.itb.ac.id}}}
\date{}

\begin{document}

\maketitle

\begin{abstract}
Alahmadi et al. ["Twisted centralizer codes", \emph{Linear Algebra and its Applications} {\bf 524} (2017) 235-249.]
introduced the notion of twisted centralizer codes, $\mathcal{C}_{\fq}(A,\gamma),$ defined as
\[
\mathcal{C}_{\fq}(A,\gamma)=\lbrace X\in\Fnn:~\ AX=\gamma XA\rbrace,
\]
for $A \in \fq^{n \times n},$ and $\gamma \in \fq.$
Moreover, Alahmadi et al. ["On the dimension of twisted centralizer codes", \emph{Finite Fields and Their Applications} {\bf 48} (2017) 43-59.]
also investigated the dimension of such codes and obtained upper and lower bounds for the dimension, and
the exact value of the dimension only for cyclic or diagonalizable matrices $A.$
Generalizing and sharpening Alahmadi et al.'s results, in this paper, we determine the exact value of the dimension as well as
provide an algorithm to construct an explicit basis of the codes for any given matrix $A.$
\end{abstract}
	
\section{Introduction}
In 2017, Alahmadi and his coauthors introduced a notion of twisted centralizer codes \cite{ax=axa}.
Let $A$ be an $n\times n$ matrix over a finite field $\fq:=GF(q)$ and $\gamma$ be an element in $\fq.$ The centralizer of $A$ twisted by $\gamma,$
denoted by $\mathcal{C}_{\fq}(A,\gamma),$ is defined as
\[
\mathcal{C}_{\fq}(A,\gamma):=\lbrace X\in\Fnn:~\ AX=\gamma XA\rbrace.
\]
It is easy to see that $\mathcal{C}_{\fq}(A,\gamma)$ is a linear subspace of the matrix space $\Fnn$ over $\fq,$ and hence
$\mathcal{C}_{\fq}(A,\gamma)$ is a linear code, called a twisted centralizer code, whose elements (i.e., codewords) are matrices, that can be viewed as vectors of length $n^2,$ by reading them  column-by-column.
The notion of twisted centralizer codes is a generalization of centralizer codes \cite{ax=xa}, since a centralizer code is a twisted centralizer code,
twisted by $1=\gamma \in \fq.$

Regarding twisted centralizer codes, so far, Alahmadi et al. (\cite{ax=axa},\cite{ondim}) have determined an upper and lower bound for the dimension, and obtained the exact values of the dimension
only for cyclic or diagonalizable matrices $A.$  Moreover, they \cite{ax=xa} also state, "In general determining the dimension of such a code \footnote{Namely, a twisted centralizer code with $\gamma=1.$ } given $A$
is a non-trivial problem, and we were only able to give a spectral characterization of the dimension
over an extension of the base field"\footnote{\cite{ax=xa}, pp. 76}.
Continuing and improving the results obtained by Alahmadi et al. (\cite{ax=axa},\cite{ondim}), the purpose of this paper is to provide the exact dimension of the codes, for any given matrix $A.$  Moreover, we also provide an algorithm to construct a basis of the codes explicitly.
The key idea to attack the problem is by observing twisted centralizer codes from the viewpoint of module theory, namely
by transforming a twisted centralizer code into a polynomial module of homomorphisms. By taking this viewpoint, we solve the "non-trivial problem"
in a more general setting, namely for arbitrary given $\gamma \in \fq,$ in an elementary way.
Furthermore, and in contrast to Alahmadi et al. approach, the investigation  of the proposed results
keeps the underlying  field being the base one; that is,  lifting the
problem over a field extension is not necessary.

The organization of the paper is as follows.  In Section 2 we provide basic facts on some structural aspects of
 polynomial modules of homomorphisms.  The main result is provided in Section 3.  The paper is ended by concluding remarks in Section 4.
We follow \cite{huffman} and \cite {modul} for undefined terms in coding theory and module theory, respectively.
	
\section{ {$\fq[x]$-module of homomorphisms}: basic facts}\label{basic facts}

From now on  $\fq[x]$ is the polynomial ring over the finite field $\fq$ of order $q$.
In this section, we provide a decomposition of an $\fq[x]$-module of homomorphisms. We do such a thing,
since we can consider twisted centralizer codes
 as $\fq[x]$-modules of homomorphisms, as we show in the next section (see Section 3).
Our decomposition will make it easy to study the structure of twisted centralizer codes.

Projection and injection homomorphisms are our main decomposition tool.
We will use a slightly abused notations of projection and injection. For an $\fq[x]$-module
with direct decomposition $M = N \oplus K$, the projection $\rho$ from $M$ on submodule $N$ along the submodule $K$,
$\rho(x+y) = x$ for all $x \in N, y \in K$, can be considered as having the codomain $N$ or $M$.
In a similar way, an injection $\iota$ from $N$ to $M$,  $\iota(x) =x$ for all $x \in N$,  can be considered as having the domain $N$ or
$M$ with $\iota(y) =0$ for all $y \in K$.

We notice that some facts are well known, but we mention here for the reader's convenience.
%
%
% \subsection{$\fq[x]$-module of homomorphisms}
%
Let $N$ and $M$ be $\fq[x]$-modules. Let ${\rm Hom}_{\fq[x]}(N,M)$ denote the set of  %$\lbrace\theta:N\to M\mid~ \theta \text{ is an $R$-module homomorphism}\rbrace.$
$\fq[x]$-module homomorphisms from $N$ to $M.$  It is clear that ${\rm Hom}_{\fq[x]}(N,M)$ is also an $\fq[x]$-module
 and we call it $\fq[x]$-module of homomorphisms (from $N$ to $M$).

Let $ A \in \fq^{n \times n}$. Let us define an action
\[
f(x)\cdot u:=f(A)u, \quad \hbox{ for every} \quad  f(x)\in \fq[x] \quad
\hbox{and} \quad  u\in \fq^{n\times m}.
\]
By this action, $\fq^{n\times m}$ can be regarded as an $\fq[x]$-module, which we call an \fxmod induced by $A$.
Alahmadi et al.  employed properties of $\fq^n$ as an $\fq[x]$-module to decompose a twisted centralizer code
 $\mathcal{C}_{\fq}(A,\gamma)$.
Further,  they noted  that  a centralizer code $\mathcal{C}_{\fq}(A,1)$ is an $\fq$-algebra and a twisted centralizer code
 $\mathcal{C}_{\fq}(A,\gamma)$ is an $\mathcal{C}_{\fq}(A,1)$-module
(see \cite[pp. 44]{ondim}).
In Section 3, with the right choice of an action we will show that  the
twisted centralizer code $\mathcal{C}_{\fq}(A,\gamma)$
is an $\fq[x]$-module of
homomorphisms.
This fact gives us  materials to study the structures of twisted centralizer codes from  module theory's viewpoint.

It is well known that the polynomial ring $\fq[x]$ is a principal ideal domain and every finitely generated torsion module over a principal ideal domain can be decomposed into its primary submodules. Further, any
finitely generated primary module over a principal ideal domain can be
decomposed as a direct sum of cyclic submodules.
 Hence, we will use those facts and the following two lemmas to derive the decomposition
  of an $\fq[x]$-module of homomorphisms between two finitely generated torsion modules.
  The first lemma is a fact concerning a decomposition of modules.
  The proof can be directly derived  from the decomposition form of both modules.

\begin{lemma}\label{dekomposisi juml langsung}
Let $N$, $M$ be two $\fq[x]$-modules having   decompositions
\[ N=N_{1}\oplus\cdots\oplus N_{n}, \quad \hbox{and}\]
\[ M=M_{1}\oplus\cdots\oplus M_{m}.\]
 Let $\rho_{i}$ be the projection from $N$ on $N_{{i}}$ along  $\oplus_{\nu =1, \nu \ne i}^n N_\nu$ for
  $i = 1, 2, \dots, n$, and $\iota_{j}$ be the injection from $M_{{j}}$ to $M$ for $j =1, 2, \dots, m$. Then
$\homofqx(N,M)=\bigoplus_{i=1}^n\bigoplus_{j=1}^m\iota_{j}\homofqx(N_{i},M_{j})\rho_{i}.$
\end{lemma}

%Now, the problem becomes simpler by transforming it into studying homomorphisms collection between primary submodules. First, %we observe homomorphisms collection between primary submodules where their orders are relatively prime.
The second lemma is a fact concerning the order of modules. For  any finitely generated torsion  $\fq[x]$-module
$M$, we denote $o(M)$ as an order of $M$.

\begin{lemma}\label{hom dg orde relatif prima}
	Let $N$ and $M$ be torsion $\fq[x]$-modules where $\fpb(o(N),o(M))=1$.
	Then
\[ \homofqx(N,M)=\{ 0\}. \]
\end{lemma}
\bukti	Let $o(N)=h$ and $o(M)=g$.
Let $\theta\in\homofqx(N,M)$ and $u\in N$.
Because $\fpb(h,g)=1$, so there exist  $a,b\in \fq[x]$ such that $ah+bg=1$.
Consider that $\theta(u)=\theta(1\cdot u)=\theta((ah+bg)\cdot u)=\theta((ah)\cdot u)+bg\cdot\theta(u)$.
Because $h=o(N)$, so $h\cdot u=0$.
Because $g=o(M)$, so $g\cdot\theta(u)=0$.
This implies $\theta(u)=0$ for every $u\in N$.
Hence $\theta=0$.
Therefore $\homofqx(N,M)=\{0\}$.\QED

As mentioned before, any finitely generated $\fq[x]$-module can be decomposed as the direct sum of its primary submodules
with order prime/irreducible power. Combining this fact and the above two lemmas we obtain the following corollary.
%Particularly, since we have $\homofqx(N_{p_i},M_{q_j})$ when ${\rm gcd}(p_i, q_j) =1$.

\begin{akibat}\label{c irisan faktor is irreducible}
Let $N$ be an \fxmod having primary decomposition
\[
N=N_{p_1}\oplus\cdots\oplus N_{p_n}
\]
with order $o(N_{p_i}) = p_i^{\nu_i}$ for some  $p_i, i =  1, \ldots, n$
different irreducible monic elements in $\fq[x]$ and $M$ be an \fxmod having primary decomposition
\[
M=M_{r_1}\oplus\cdots\oplus M_{r_m}
\]
with order $o(M_{r_j}) = r_j^{\mu_j}$ for some distinct irreducible monic elements $r_j$ in $\fq[x],$ $j = 1, \ldots, m$.
Let $\rho_{p_i}$ be the projection from $N$ on $N_{p_i}$ along
$\oplus_{\tau=1, \tau \ne i}^n N_{p_\tau}$ and $\iota_{r_j}$ be the injection from $M_{r_j}$ to $M.$ It follows then
\[ \homofqx(N,M)=\bigoplus_{i=1}^k\iota_{s_i}\homofqx(N_{s_i},M_{s_i})\rho_{s_i}
\]
where $\lbrace s_1,\cdots,s_k\rbrace=\{p_1,\cdots,p_n \}\cap\{r_1,\cdots,r_m \}$.
	
\end{akibat}
The above corollary uses the fact that  $\homofqx(N_{p_i},M_{r_j})= \{0\}$ when ${\rm gcd}(p_i, r_j) =1$.

Next, let us look at the case when the modules $N$ and $M$ are primary modules whose order are not relatively prime. Regarding the primary cyclic decomposition
(see, e.g., \cite{modul}), we know
that primary submodules can be decomposed as a direct sums of cyclic submodules. Combining this fact and Lemma \ref{dekomposisi juml langsung}, we obtain the following corollary.

\begin{akibat}\label{dekomposisi hom where primer sama}
Let $p$ be an irreducible element in $\fq[x]$ and $N, M$ be primary $\fq[x]$-modules
having cyclic decompositions as
\[ N=\langle u_1\rangle\oplus\cdots\oplus\langle u_k\rangle\]
 where $o(u_i)=p^{a_i}$ for every $i=1,\ldots, k, \quad a_1\geq a_2\geq\cdots\geq a_k,$ and
 \[ M=\langle v_1\rangle\oplus\cdots\oplus\langle v_\ell \rangle\]
  where $o(v_j)=p^{b_j}$ for every $j=1,\ldots,\ell $  and $b_1\geq b_2\geq\cdots\geq b_\ell.$
Let $\iota_j$ be the injection from $\langle v_j\rangle$ to $M$ for every $j= 1,\ldots, \ell$ and $\rho_i$ be the projection from $N$ on $\langle u_i\rangle$ along $ \oplus_{\nu =1, \nu \ne i }^n \langle u_\nu \rangle$
for every $i = 1,\ldots,k.$
Then $\homofqx(N,M)= \bigoplus_{i=1}^k\bigoplus_{j=1}^\ell \iota_j\homofqx(\langle u_i\rangle,\langle v_j\rangle)\rho_i.$
\end{akibat}

We now determine a generator of an $\fq[x]$-module  of homomorphisms between two cyclic $\fq[x]$-modules which is
obviously cyclic.  For that purpose
we first define the following.
For $f,g\in \fq[x],$ define $\min\lbrace f,g\rbrace:=\left\lbrace\begin{array}{cc}
f,&\text{ if }\deg f\le\deg g,\\
g, &\text{ if }\deg f>\deg g.
\end{array}\right.$

\begin{lemma}\label{ou|ov}
	Let $\langle u\rangle$ and $\langle v\rangle$ be cyclic \fxmod \!\!s where $\fpb(o(u),o(v))=\min\lbrace o(u), o(v)\rbrace$.
	Then we have
%	\begin{enumerate}
%		\item $o(\homofqx(\langle u\rangle,\langle v\rangle))=\fpb(o(u),o(v))$ and
\[
\homofqx(\langle u\rangle,\langle v\rangle)=\langle\theta\rangle,  \text{ where }\theta(u)=\dfrac{o(v)}{\fpb( o(u),o(v))}\cdot v.
\]
%	\end{enumerate}
\end{lemma}
\bukti 	
Let $\fpb(o(u),\ o(v))=d$ and $o(v)=dr.$
%\ o(\homofqx(\langle u\rangle,\langle v\rangle))=f,\ o(v)=g$, and $o(u)=h.$
%Let $d=ag+bh$ for some $a,b\in \fq[x]$
Consider $\theta\in\homofqx(\langle u\rangle,\langle v\rangle)$ which maps $u$ to $r\cdot v.$
%Because $f$ is $o(\homofqx(\langle u\rangle,\langle v\rangle))$, so $0=f\cdot\theta(u)=(fr)\cdot v.$
%Then $o(v)\mid fr,$ which implies $d\mid f.$
%Let $\varphi\in\homofqx(\langle u\rangle,\langle v\rangle).$
%Notice that $d\cdot\varphi(u)=(ag+bh)\cdot\varphi(u)=(ag)\cdot\varphi(u)+b\cdot\varphi(h\cdot u)=0.$
%Hence $f\mid d,$ and therefore $d$ is $o(\homofqx(\langle u\rangle,\langle v\rangle)).$
Let
$\varphi \in\homofqx(\langle u\rangle,\langle v\rangle),$
$\varphi(u)=p \cdot v$ for some $p\in \fq[x].$
Then $d(p\cdot v)=d\cdot\varphi(u)=0.$
This implies $o(v)\mid dp,$
and hence $r\mid p,$ which is equivalent to $p=rs$ for some $s\in \fq[x].$
Hence $\varphi(u)=(sr)\cdot v=s\cdot\theta(u)$ and $\varphi\in\langle\theta\rangle.$
Therefore $\homofqx(\langle u\rangle,\langle v\rangle)=\langle\theta\rangle.$\QED

Let us observe an example.

\begin{contoh}
Let $N=\FF_3^3$ be an $\FF_3$-module  induced by
	$\begin{bmatrix}
	0&0&0\\
	0&0&1\\
	0&0&0
	\end{bmatrix}$.
	Then its primary cyclic decomposition  is $N=\underbrace{\langle[1\ 0\ 0]^t \rangle\oplus\langle[0\ 0\ 1]^t\rangle}_{N_{(x)}}$ where $o([1\ 0\ 0]^t)=x$ and $o([0\ 0\ 1]^t)=x^2.$ Let $M=\FF_3^2$ be an $\FF_3$-module induced by
	$\begin{bmatrix}
	0&0\\
	1&0
	\end{bmatrix}$. Then its primary cyclic decomposition is $M=\underbrace{\langle[1\ 0]^t\rangle}_{M_{(x)}}$ where $o([1\ 0]^t)=x^2.$ Then \begin{align*}
\homo_{\FF_3[x]}(N,M)=&\iota_1\homo_{\FF_3[x]}(\langle [1\ 0\ 0]^t \rangle,\langle[1\ 0]^t\rangle)\rho_1\oplus\iota_1\homo_{\FF_3[x]}(\langle[0\ 0\ 1]^t\rangle,)\langle[1\ 0]^t\rangle\rho_2\\
	=&\left\langle\begin{bmatrix}
	0&0&0\\1&0&0
	\end{bmatrix}\right\rangle\oplus\left\langle\begin{bmatrix}
	0&1&0\\0&0&1
	\end{bmatrix}\right\rangle.
\end{align*} $\diamondsuit$
\end{contoh}

From the explanation above we identify that the orders of primary submodules as well as the orders of  cyclic submodules in
their decomposition play an important role on determining the structure of $\fq[x]$-module of homomorphisms.
The multiset of orders of cyclic submodules in the decomposition of primary submodules of a module $M$ is called
the elementary divisors of $M$ and denoted by ${\rm ElemDiv}(M)$ \cite{modul}.

\section{Bases for and dimensions of twisted centralizer codes}\label{Primary Cyclic Decomposition of Twisted Centralizers Code}

In this section we determine a basis for and the dimension of a twisted centralizer code. We  show first, that a twisted centralizer code is nothing but an $\fq[x]$-module of homomorphisms. As a consequence, we can decompose a twisted centralizer code as we did in the previous section.
At the end, we determine a primary cyclic decomposition of a twisted centralizer code, and then extract a basis for and the dimension of it explicitly. %Also in this section we will prove some results in references in an alternate way, like a generalization way before.
%We also show that main results in \cite{ax=xa}, \cite{ax=axa} are corollaries of our main result.

We divide it  into two cases: twisted centralizer codes twisted by $\gamma=0,$ and twisted by $\gamma\ne 0$.  To obtain the result for twisted centralizer codes twisted by $\gamma\ne 0,$ we consider first twisted polynomials and derive their properties.

\subsection{Twisted centralizer codes with $\gamma=0$}

We obtain a theorem regarding the centralizer code of $A$ which is twisted by $0,$ namely
\[ \mathcal{C}_{\fq[x]}(A,0)=\lbrace X \in \fq^{n\times n}:~ AX=0 \rbrace.\]

\begin{teorema}\label{C(A,0)}
Let a primary cyclic decomposition of $\fq^n$ as \fxmod induced by $A$ be
\[
\fq^n=\underbrace{\left[\langle v_{1,1}\rangle\oplus\cdots\oplus\langle v_{1,l(1)}\rangle  \right]}_{(\fq^n)_{p_1}}\oplus\cdots\oplus \underbrace{\left[\langle v_{m,1}\rangle\oplus\cdots\oplus\langle v_{m,l(m)}\rangle  \right]}_{(\fq^n)_{p_m}}
\]
where $p_1=x$ and $o(v_{k,j})=p_k^{s(k,j)}$ for every $k=1,\ldots,m$ and $j=1,\ldots,l(k).$ Then
\[
\mathcal{C}_{\fq[x]}(A,0)= \bigoplus_{i=1}^n\bigoplus_{j=1}^{l(1)}\langle E^i_{x^{s(1,j)-1}v_{1,j}}\rangle
\]
where $E^i_{x^{s(1,j)-1}v_{1,j}}$ is in $\fq^{n\times n}$ with zero columns except the $i$-th column which is equal to $x^{s(1,j)-1}\cdot v_{1,j}$.
\end{teorema}

\bukti
Obviously, $ \bigoplus_{i=1}^n\bigoplus_{j=1}^{l(1)}\langle E^i_{x^{s(1,j)-1}v_{1,j}}\rangle\subseteq \mathcal{C}_{\fq[x]}(A,0).$
Now let $X\in \mathcal{C}_{\fq[x]}(A,0).$
Because $\fq^{n\times n}= \bigoplus_{i=1}^n\bigoplus_{k=1}^m\bigoplus_{j=1}^{l(k)} \langle E^i_{v_{k,j}}\rangle,$
so $X$ can be stated as $X= \sum_{i=1}^n\sum_{k=1}^m\sum_{j=1}^{l(k)}f_{i,k,j}E^i_{v_{k,j}}$ where $f_{i,k,j}\in \fq[x].$
Since $X\in\mathcal{C}_{\fq[x]}(A,0)$, we get
$0=AX= \sum_{i=1}^n\sum_{k=1}^m\sum_{j=1}^{l(k)}xf_{i,k,j}E^i_{v_{k,j}}.$
This implies $p_k^{s(k,j)}\mid xf_{i,k,j}.$
Because $\fpb(x,p_k)=1$ for $k\neq 1,$ so $p_k^{s(k,j)}\mid f_{i,k,j}.$
Hence $f_{i,k,j}E^i_{v_{k,j}}=0,$ for $k\neq 1.$
For $k=1,$ we get $x^{s(1,j)}\mid xf_{i,k,j}.$
Then $f_{i,k,j}\in \langle x^{s(1,j)-1}\rangle.$
This implies $X\in  \bigoplus_{i=1}^n\bigoplus_{j=1}^{l(1)}\langle E^i_{x^{s(1,j)-1}v_{1,j}}\rangle.$
Hence $\mathcal{C}_{\fq[x]}(A,0)\subseteq  \bigoplus_{i=1}^n\bigoplus_{j=1}^{l(1)}\langle E^i_{x^{s(1,j)-1}v_{1,j}}\rangle$. Therefore $\mathcal{C}_{\fq[x]}(A,0)=\bigoplus_{i=1}^n\bigoplus_{j=1}^{l(1)}\langle E^i_{x^{s(1,j)-1}v_{1,j}}\rangle$.\QED

The above theorem gives a structure of twisted centralizer codes of $A$ twisted by $0$ as a module. Then as a vector space, the dimension of a centralizer code of $A$ twisted by $0$ can be directly determined.

\begin{akibat}\label{L2.1-ondim}
	Let $A\in\Fnn$ has nullity $k_0.$ Then $\dim\mathcal{C}_{\fq[x]}(A,0)=k_0 n.$
\end{akibat}

\bukti
Let a primary cyclic decomposition of $\fq^n$ be given by
\[
\fq^n=\underbrace{\left[\langle v_{1,1}\rangle\oplus\cdots\oplus\langle v_{1,l(1)}\rangle  \right]}_{{(\fq^n)}_{p_1}}\oplus\cdots\oplus \underbrace{\left[\langle v_{m,1}\rangle\oplus\cdots\oplus\langle v_{m,l(m)}\rangle  \right]}_{{(\fq^n)}_{p_m}},
\]
where $p_1=x$ and $o(v_{k,j})=p_k^{s(k,j)}$ for every $k=1,\ldots,m$ and $j=1,\cdots,l(k).$
By Theorem  \ref{C(A,0)}, $\ker(A)= \bigoplus^{l(1)}_{j=1}\langle x^{s(1,j)-1}v_{1,j}\rangle.$
Then $k_0=l(1).$
Again, by Theorem \ref{C(A,0)}, $\mathcal{C}_{\fq[x]}(A,0)= \bigoplus_{i=1}^n\bigoplus_{j=1}^{l(1)}\langle E^i_{x^{s(1,j)-1}v_{1,j}}\rangle$.
Then $\dim\mathcal{C}_{\fq[x]}(A,0)=nl(1)=nk_0$.
\QED

\subsection{Twisted polynomial}

We consider twisted polynomials and derive their properties.
The following definition is adapted from twisted characteristic polynomial in  \cite{ondim}.

\begin{definisi}%[Alahmadi, dkk\cite{ondim}]
Let $0\neq \gamma\in \fq,f(x)\in \fq[x],$ and $\deg(f)$ be the degree of $f.$
The $\gamma$-twisted polynomial of $f(x)$, denote by $\fgamma(x),$ is defined as
$\fgamma(x):=\gamma^{\deg(f)}f(x/\gamma)$. %We call $\fgamma(x)$ and is called by \textbf{$f(x)$ which is twisted by $\gamma$}.
\end{definisi}

\begin{catatan}
Observe that $\fsepergamma(x)=\gamma^{-\deg(f)}f(\gamma x)$. Moreover, $\fsepergamma$ and $\fgamma$ are monic if and only if $f$ is monic.
\end{catatan}

Since $\deg(f(x))=\deg(f(\gamma x))=\deg(f(x/\gamma))$, we also have
\begin{equation}\label{derajat from f,fgamma,fsepergamma}
\deg(f)=\deg(\fgamma)=\deg(\fsepergamma).
\end{equation}
%
%\begin{proposisi}\label{derajat from f,fgamma,fsepergamma} Degree of $f,\fgamma,$ and $\fsepergamma$ is equal or
%	$\deg(f)=\deg(\fgamma)=\deg(\fsepergamma)$.
%\end{proposisi}
%\bukti	
%We can obtain the fact above because $\deg(f(x))=\deg(f(\gamma x))=\deg(f(x/\gamma))$.\QED

\begin{proposisi}\label{f=gh-fgamma=ggamma.hgamma}
Let $f(x)\in \fq[x]$. Then the following statements are equivalent.
\begin{itemize}
\item[(1)] $f(x)=g(x)h(x).$
\item[(2)] $\fgamma(x)=g^{\gamma}(x)h^{\gamma}(x).$
\item[(3)] $\fsepergamma(x)=g^{1/\gamma}(x)h^{1/\gamma}(x).$
\end{itemize}		
\end{proposisi}
\bukti

((1) $\Rightarrow$ (2)) $\fgamma(x)=\gamma^{\deg(f)}f(x/\gamma)=\gamma^{\deg(gh)}g(x/\gamma)h(x/\gamma)=g^{\gamma}(x)h^{\gamma}(x)$.

((2) $\Rightarrow$ (3)) $\fsepergamma(x)=\gamma^{-2\deg(f)}\fgamma(\gamma^2 x)=\gamma^{-2\deg(g)}g^{\gamma}(\gamma^2x)\gamma^{-2\deg(h)}h^{\gamma}(\gamma^2x)= g^{1/\gamma}(x)h^{1/\gamma}(x)$.

((3) $\Rightarrow$ (1)) $f(x)=\gamma^{\deg(f)}\fsepergamma(x/\gamma)=\gamma^{\deg(gh)}g^{1/\gamma}(x/\gamma)h^{1/\gamma}(x/\gamma)=g(x)h(x)$.\QED

As a direct result, we obtain the following Lemma.

\begin{lemma}\label{tak tereduksinya f,fgamma,fsepergama}
Let $f(x)\in \fq[x]$. Then the following statements are  equivalent.
\begin{itemize}
\item[(1)] $f(x)$ is irreducible.
\item[(2)] $\fgamma(x)$ is irreducible.
\item[(3)] $\fsepergamma(x)$ is irreducible.
\end{itemize}
\end{lemma}

By Equation (\ref{derajat from f,fgamma,fsepergamma}) and Proposition \ref{f=gh-fgamma=ggamma.hgamma} we obtain the following lemma.

\begin{lemma}
Let $f,g\in \fq[x]$. Then the following statements are equivalent.
\begin{itemize}
\item[(1)] $\fpb(f,g)=h.$
\item[(2)] $\fpb(\fgamma,g^{\gamma})=h^{\gamma}.$
\item[(3)] $\fpb(\fsepergamma,g^{1/\gamma})=h^{1/\gamma}.$
\end{itemize}	
\end{lemma}

Let $A\in \Fnn.$ Observe that $f(A)=\gamma^{-\deg(f)}\fgamma(\gamma A)$.
This implies $f(A)=0$ if and only if $\fgamma(\gamma A)=0$. Then we have the lemma below.

\begin{lemma}\label{mA iff myA}
Let $m_A(x)$ be the minimal polynomial of $A.$ Then the minimal polynomial of $\gamma A$ is	$m_A^\gamma(x).$
\end{lemma}

\subsection{Twisted centralizer codes with $\gamma\neq 0$}

In this section we determine a basis for and the dimension of
a twisted centralizer code by examining relations between twisted centralizer codes and collection of module homomorphisms.
%We can remember that centralizers code is equal to endomorphism collection.

Let $S\in\Fnn.$ Let $[S]\fq^n$ denote $\fq^n$ as the \fxmod  is induced by $S.$
The following is the key theorem that opens a way to utilize the structure of $\fq[x]$-module of homomorphisms
discussed in Section 2 to obtain a basis of a twisted centralizer code and hence its dimension.

\begin{teorema}\label{kode pemusat terpelintir=hom}
	Let $A\in \Fnn$ and $\gamma\neq 0$.
	Then $\mathcal{C}(A,\gamma)=\homofqx(\pengN, \pengM)$.
\end{teorema}

\bukti	
Let $X\in\mathcal{C}(A,\gamma)$.
Let $f(x),g(x)\in F[x]$, and $u,v\in [\gamma A]\fq^n$.
Then
\[
\begin{aligned}
X(f(x)\cdot u+g(x)\cdot v) & = X(f(\gamma A)u+g(\gamma A)v)\\
                           & = Xf(\gamma A)u+Xg(\gamma A)v\\
                           & = f(A)Xu+g(A)Xv\\
                           & = f(x)\cdot (Xu)+g(x)\cdot (Xv).
\end{aligned}
\]
Hence $X\in \homofqx(\pengN,\pengM)$. It follows that  $\mathcal{C}(A,\gamma)\subseteq\homofqx(\pengN,\pengM)$.
\\
Let $X\in\homofqx(\pengN,\pengM)$.
Then % Let $u\in \fq^n$.
\[
%\begin{array}{rcl}
\gamma XAu=X(\gamma Au)=X(x\cdot u)=x\cdot (Xu)=A(Xu)=AXu \quad
 \hbox{ for all} \quad u\in [\gamma A]\fq^n.\]
Thus $\gamma XA=AX$ which implies $X\in\mathcal{C}(A,\gamma)$.
Hence $\homofqx(\pengN,\pengM)\subseteq\mathcal{C}(A,\gamma)$.	We conclude that  $\mathcal{C}(A,\gamma)=\homofqx(\pengN,\pengM)$.\QED

Note that for the case $\gamma =1$, Theorem \ref{kode pemusat terpelintir=hom} says that the centralizer code $\mathcal{C}(A,1)$ is equal to
 the $\fq[x]$-module of  endomorphisms on $[A]\fq^n$.
%Next we provide a decomposition of it.

In order to apply the results in Section 2 to a twisted centralizer code $\mathcal{C}(A,\gamma)$, we need to
identify the primary and cyclic decomposition of $[A]\fq^n$ and $[\gamma A]\fq^n$. The following
two corollaries show the relation between orders of $[A]\fq^n$ and $[\gamma A]\fq^n$,
between the primary and cyclic decompositions of $[A]\fq^n$ and $[\gamma A]\fq^n$,
as well as between ${\rm ElemDiv}([A]\fq^n)$ and ${\rm ElemDiv}([\gamma A]\fq^n)$.

%\begin{teorema}
%Let a primary cyclic decomposition of $\fq^n$ as \fxmod be \[\fq^n=\underbrace{\left[\langle %v_{1,1}\rangle\oplus\cdots\oplus\langle v_{1,l(1)}\rangle  \right]}_{(\fq^n)_{p_1}}\oplus\cdots\oplus %\underbrace{\left[\langle v_{m,1}\rangle\oplus\cdots\oplus\langle v_{m,l(m)}\rangle  \right]}_{(\fq^n)_{p_m}}  \] where $p_1,\cdots,p_m$ is different irreducible monic elements in $\fq[x]$ and $o(v_{k,j})=p_k^{s(k,j)}$ where $s_k=s(k,1)\geq %s(k,2)\geq\cdots\geq s(k,l(k))$ for  every $k=1,\cdots,m$.
%	Then $$\ndo_{F[x]}\fq^n=\ndo_{F[x]}(\fq^n)_{p_1}\oplus\cdots\oplus\ndo_{F[x]}(\fq^n)_{p_m}$$
%	where $o(\ndo_{F[x]}(\fq^n)_{p_i})=p_i^{s_i}$ for $i=1,\cdots,m$.
%\end{teorema}
%\bukti
%From Lemma \ref{hom dg orde relatif prima} we get for every $\theta:F^n_{p_i}\to\bigoplus_{j\neq i}(\fq^n)_{p_j}$ hold %$\theta=0$.
%Consider that $o(\ndo_{F[x]}(\fq^n)_{p_i})\mid p_i^{s_i}$.
%This implies  $$\ndo_{F[x]}\fq^n=\ndo_{F[x]}(\fq^n)_{p_1}\oplus\cdots\oplus\ndo_{F[x]}(\fq^n)_{p_m}.$$
%Consider a \fxmod homomorphism $\phi:\langle v_{i,1}\rangle\to \langle v_{i,1}\rangle$ where $\phi(v_{i,1})=v_{i,1}$.
%This homomorphism has an order $p_i^{s(i,1)}=p_i^{s_i}$. Then $p_i^{s_i}\mid o(\ndo_{F[x]}F^n_{p_i})$. This implies %$o(\ndo_{F[x]}F^n_{p_i})=p_i^{s_i}$, and it holds for every $i=1,\cdots,m.$\QED

\begin{akibat}\label{primary cyclic decomposition N}
	Let $A\in \Fnn$ and $\gamma\ne 0$.
	Let $\pengM$ has an order $p_1^{s_1}p_2^{s_2}\cdots p_m^{s_m}$ where $p_1,\cdots,p_m$ are distinct irreducible monic elements in $\fq[x]$ and $s_1,\cdots,s_m\in \mathbb{N}$.
	Then $\pengN$ has an order $(\pgamma_1)^{s_1} \cdot (\pgamma_2)^{s_2}\cdots (\pgamma_m)^{s_m}$
\end{akibat}
\bukti
From Lemma \ref{mA iff myA}, we obtain an order of $\pengN$ is
\[
m_{\gamma A}(x)=m^{\gamma}_A(x)=\break(p_1^{s_1}p_2^{s_2}\cdots p_m^{s_m})^{\gamma}.
\]
Using Corollary \ref{f=gh-fgamma=ggamma.hgamma}, we conclude that
\[o(\pengN)=m_{\gamma A}(x)=(\pgamma_1)^{s_1}(\pgamma_2)^{s_2}\cdots (\pgamma_m)^{s_m}.\] \QED

Now, let $v \in \fq^n$ and $f(x) \in \fq[x]$. Considering $\fq^n$ as the   $\pengM-$module, the action of
$f(x)$ on $v$ produces
\[
f(x) \cdot v = f(A) v.
\]
  Meanwhile, considering $\fq^n$  as the $\pengN$ module,
the action of
$f^\gamma (x)$ on $v$ produces
\begin{equation}\label{tw1}
 f^\gamma (x) \cdot v =\gamma^{\deg (f)} f\left(\frac{x}{\gamma}\right) \cdot v
= \gamma^{\deg (f)}f\left(\frac{\gamma A}{\gamma}\right)v =\gamma^{\deg(f)} f(A) v.
\end{equation}
Thus, the action of $f(x)$ on  $ v  $ in the context of $\pengM-$ module will produce a scalar multiple of
the action of $f^{\gamma}(x)$ on $v$ in the context of $\pengN-$ module. By utilizing this property we obtain
the following.

\begin{akibat}\label{decomposition N's primary}
	If one of the primary submodules of $\pengM$ is
$(\pengM)_{p_k}$ % is the primary submodule of $\pengM$
with a cyclic decomposition \[(\pengM)_{p_k}=\langle v_{k,1}\rangle\oplus\cdots\oplus\langle v_{k,l(k)}\rangle \] where $o(v_{k,j})=(p_k)^{s(k,j)}$ and $s_k=s(k,1)\geq s(k,2)\geq\cdots\geq s(k,l(k))$, then
one of primary submodules of $\pengN$ is $(\pengN)_{\pgamma_k}= (\pengM)_{p_k}
$ with a cyclic decomposition
%	Then
 \[(\pengN)_{\pgamma_k}=\langle v_{k,1}\rangle\oplus\cdots\oplus\langle v_{k,l(k)}\rangle \] where $o(v_{k,j})=(\pgamma_k)^{s(k,j)}$ and $s_k=s(k,1)\geq s(k,2)\geq\cdots\geq s(k,l(k))$.	
\end{akibat}

\bukti
Let $C= \langle v_{k,j}\rangle$ be a cyclic submodule of $\pengM$ with   $o(v_{k,j})=(p_k)^{s(k,j)}$. It suffices to show
that $C$ is also the cyclic submodule of $\pengN$ generated by $v_{k,j}$ with   $o(v_{k,j})=(\pgamma_k)^{s(k,j)}$.
Let $f(x) \in \fq[x]$. Then there exists $g(x) \in \fq[x]$ such that $  f(x) = g^{\gamma}(x)$.
 Considering (\ref{tw1}), the action of $f(x)$ on $v_{k,j}$, in view of $\pengN$ module, implies
 \[
 f(x) \cdot v_{k,j} = \gamma^{\deg(g)} g(A)v_{k,j} \in C.
 \]
  Conversely, by applying (\ref{tw1}), any element of $C$ is an element of the cyclic submodule of $\pengN$ generated by $v_{k,j}$. Thus $C$ is the cyclic submodule of $\pengN$ generated by
$v_{k,j}$. Applying (\ref{tw1}) once more, we obtain the order of $v_{k,j}$ as an element  of $\pengN$
is $o(v_{k,j})=(\pgamma_k)^{s(k,j)}$.

\QED

The above corollary implies that the primary and cyclic decompositions of $[A]\fq^n$ and $[\gamma A]\fq^n$ are
the same, even though their orders are twisted. Moreover,
\begin{equation}\label{elementary divisor gammaA}
{\rm ElemDiv}([\gamma A]\fq^n) = \{ f^\gamma(x) \ : \ f(x) \in {\rm ElemDiv}([A]\fq^n)\}.
\end{equation}
As a direct implication of Theorem \ref{kode pemusat terpelintir=hom} and Lemma \ref{dekomposisi juml langsung},
%\label{dekomposisi hom antar primary},
 we have the following corollary.

\begin{akibat}\label{akibat-dimensi} %\label{kode=sum primer}
\begin{itemize}
\item[(1)]	Let $\rho_{\pgamma_i}$ be the projection from $\pengN$ to $(\pengN)_{\pgamma_{i}}$ and $\iota_{p_k}$ be the  injection from $(\pengM)_{p_{k}}$ to $\pengM$. Then
\[ \mathcal{C}(A,\gamma)=\homofqx(\pengN,\pengM)=\bigoplus_{i=1}^m\bigoplus_{k=1}^m\iota_{p_k}\homofqx((\pengN)_{\pgamma_i},(\pengM)_{p_k})\rho_{\pgamma_i}.
\]
%\end{akibat}

%\begin{akibat}\label{hom primer=sum hom siklik}
\item[(2)] 	Let $\rho_{(\pgamma_i,c)}$ be the projection from $(\pengN)_{\pgamma_{i}}$ to $\langle v_{i,c}\rangle$ and $\iota_{(p_k,d)}$ be the injection from $\langle v_{k,d}\rangle$ to $(\pengM)_{p_{k}}$. Then
	 $$\homofqx((\pengN)_{\pgamma_i},(\pengM)_{p_k})=\bigoplus_{c=1}^{l(i)}\bigoplus_{d=1}^{l(k)}\iota_{(p_k,d)}\homo_{\fq[x]}(\langle v_{i,c}\rangle,\langle v_{k,d}\rangle)\rho_{(\pgamma_i,c)}.$$
%\end{akibat}
%\bukti
%Corollary from Lemma \ref{jumlah langsung hom submodul proyeksi injeksi}.\QED
%\begin{akibat}\label{dim siklik}
\item[(3)]	As a vector space over $\fq$,
	\[
	\dim\homo_{\fq[x]}(\langle v_{i,c}\rangle,\langle v_{k,d}\rangle)
= \deg\fpb(\pgamma_i,p_k)\min\lbrace s(i,c),s(k,d)\rbrace.
	\]
%\end{akibat}
%\begin{akibat}\label{dim primary}
\item[(4)]	As a vector space over $\fq$,
	\begin{align*}
	\dim\homofqx((\pengN)_{\pgamma_i},(\pengM)_{p_k})=&\sum_{c=1}^{l(i)}\sum_{d=1}^{l(k)}\dim\homo_{\fq[x]}(\langle v_{i,c}\rangle,\langle v_{k,d}\rangle).
	\end{align*}
\end{itemize}
\end{akibat}

% Let ElemDiv$(M)$ denote a \textit{multiset} of elementary divisors from module $M.$
Statement in the Corollary \ref{akibat-dimensi} (3) can be considered as the degree of
 the greatest common divisor of orders of  the involved cyclic submodules. Combining this observation,
  Corollary \ref{akibat-dimensi} (2), and Equation (\ref{elementary divisor gammaA})
we obtain the following theorem.

\begin{teorema}\label{dimC=fpb elemdiv}
	As a vector space over $\fq,$
\begin{align*}
	\dim\mathcal{C}(A,\gamma)=\sum_{f,g\in\text{ElemDiv}(\pengM)}\deg\fpb(f^\gamma
	,g).
	\end{align*}
\end{teorema}

\begin{contoh}
	Let $F=\FF_5$ and $A\in F^{11\times 11}$ having primary cyclic decomposition \[[A]F^{11}=\underbrace{\left[\langle v_{1,1}\rangle\oplus\langle v_{1,2}\rangle\right]}_{[A]F^4_{(x)}} \oplus\underbrace{\left[\langle v_{2,1}\rangle\oplus\langle v_{2,2}\rangle\right]}_{[A]F^4_{(x^2+x+1)}}\oplus\underbrace{\left[\langle v_{3,1}\rangle\right]}_{[A]F^4_{(x^2+2x+4)}}  \]
	where $o(v_{1,1})=x^2,\  o(v_{1,2})=x,\ o(v_{2,1})=(x^2+x+1)^2,\ o(v_{2,1})=x^2+x+1,$ and $o(v_{3,1})=x^2+2x+4$.
Let us calculate the $\dim\mathcal{C}(A,2)$. First, we have
\[\text{ElemDiv}(\pengM) = \lbrace x^2,\ x,\ (x^2+x+1)^2,\ x^2+x+1,\ x^2+2x+4\rbrace, \text{and}\]
\[\text{ElemDiv}([2A]F^{11})= \lbrace x^2,\ x,\ (x^2+2x+4)^2,\ x^2+2x+4,\ x^2+4x+1\rbrace.\]
 Then
applying Theorem \ref{dimC=fpb elemdiv} we obtain
 \[
	\dim\mathcal{C}(A,2) =2+1+1+1+2+2 = 9.\]
	$\diamondsuit$
\end{contoh}

Let us consider  the case where $A$ is a cyclic matrix. In this case, all the associated primary submodules are cyclic and
 its minimal polynomial is equal to the characteristic polynomial.
 Further, the collection of its elementary divisors is in fact a set (not a multiset) in which every two elements are relatively prime.
  Similar facts are obtained for the
 twisted matrix $\gamma A$. Hence, if $g$ is an elementary divisor of $A$, then there is at most
 one elementary divisor $f$ such that $f^\gamma$ and $g$ are not relatively prime.
 As a result
 \[ \prod_{f \in \text{ElemDiv}(\pengM)}\text{gcd}(f^\gamma, g) = \gcd(m_A(x)^\gamma, g).
 \]
 And so
 \[ \prod_{f, g \in \text{ElemDiv}(\pengM)}\text{gcd}(f^\gamma, g) =\prod_{g \in \text{ElemDiv}(\pengM)} \gcd(m_A(x)^\gamma, g) = \text{gcd}(m_A^\gamma(x), m_A(x)).
 \]
  Applying Theorem \ref{dimC=fpb elemdiv}, we can identify	
 \[
\dim\mathcal{C}(A,\gamma) \sum_{f,g \in\text{ElemDiv}(\pengM)}\deg\fpb(f^\gamma,g)= \deg\gcd(m^\gamma_A,m_A).
\]

 Thus we obtain \cite[Theorem 4.2]{ondim}, i.e., if $A$ is cyclic, the dimension of $\mathcal{C}(A,\gamma)$ is equal to the degree of greatest common divisors of the characteristic polynomials of $A$ and $\gamma A$.

As a result of the above discussions and results,
 we obtain a procedure to construct a basis of a twisted centralizer code
$\mathcal{C}(A,\gamma).$
For this purpose,   it is necessary to translate Lemma \ref{ou|ov} in a vector space context \cite[p.166]{modul}.

\begin{teorema}\label{basis siklik}
As a vector space over $\fq,$ a basis for $\homo_{\fq[x]}(\langle v_{i,c}\rangle,\langle v_{k,d}\rangle)\ne \lbrace0\rbrace$ is
\[
\lbrace \theta,A\theta,A^2\theta,\cdots,A^{\deg p_k\min\lbrace s(i,c),s(k,d)\rbrace-1}\theta\rbrace,
\]
where $\theta(v_{i,c})=p_k^{\max\lbrace s(k,d)-s(i,c),0\rbrace  }(A)v_{k,d}.$
\end{teorema}

\begin{alg}
The following algorithm produces a basis for $\mathcal{C}(A,\gamma).$
	\begin{enumerate}
		\item
		Given $A\in\fq^n$ and $\gamma\in\fq, \gamma \ne 0$.
		\item
		Construct a primary cyclic decomposition of $\pengM$. Let it be \[\pengM=\underbrace{\left[\langle v_{1,1}\rangle\oplus\cdots\oplus\langle v_{1,l(1)}\rangle  \right]}_{(\pengM)_{p_1}}\oplus\cdots\oplus \underbrace{\left[\langle v_{m,1}\rangle\oplus\cdots\oplus\langle v_{m,l(m)}\rangle  \right]}_{(\pengM)_{p_m}}  \] where $p_1,\cdots,p_m$ are different irreducible monic elements in $\fq[x]$ and $o(v_{k,j})=p_k^{s(k,j)}$ where $s_k=s(k,1)\geq s(k,2)\geq\cdots\geq s(k,l(k))$ for  every $k=1,\cdots,m$.
		% \item
\\ We obtain a	
		 primary cyclic decomposition of $\pengN$ %. It is
\[\pengN=\underbrace{\left[\langle v_{1,1}\rangle\oplus\cdots\oplus\langle v_{1,l(1)}\rangle  \right]}_{(\pengN)_{p_1^\gamma}}\oplus\cdots\oplus \underbrace{\left[\langle v_{m,1}\rangle\oplus\cdots\oplus\langle v_{m,l(m)}\rangle  \right]}_{(\pengN)_{p_m^\gamma}}  \] where $p_1,\cdots,p_m$ are different irreducible monic elements in $\fq[x]$ in  point 2. above and $o(v_{k,j})=(p_k^\gamma)^{s(k,j)}$ where $s_k=s(k,1)\geq s(k,2)\geq\cdots\geq s(k,l(k))$ for  every $k=1,\cdots,m$.\\ See Corollary \ref{primary cyclic decomposition N} and Corollary \ref{decomposition N's primary}.
		\item
		Set $TCCB=\emptyset$. ($TCCB$ stands for Twisted Centralizers Code's Bases)
		\item
		For $i$ from 1 to $m$ and $k$ from 1 to $m$.
		\begin{itemize}
			\item
			Set $HPB=\emptyset$. ($HPB$ stands for Hom Primary's Bases)
			\item
			If $p_i^\gamma= p_k$.\\ For $c$ from 1 to $l(i)$ and $d$ from $1$ to $l(k)$.
			\begin{itemize}
				\item
				$HPB=HPB\cup\left\lbrace\vartheta,A\vartheta,\cdots,A^{\deg p_k\min\lbrace s(i,c),s(k,d)\rbrace-1}\vartheta\right\rbrace$ where $\vartheta=\iota_{p_k}\iota_{(p_k,d)}\theta\rho_{(\pgamma_i,c)}\rho_{\pgamma_i},\\ \theta(v_{i,c})=p_k^{\max\lbrace s(k,d)-s(i,c),0\rbrace}(A)v_{k,d}\text{ and }\theta\in\homofqx(\langle v_{i,c}\rangle,\langle v_{k,d}\rangle).$\\	See Corollary \ref{akibat-dimensi} and Theorem \ref{basis siklik}.
				
			\end{itemize}
			\item Else, continue. \\ See Lemma \ref{hom dg orde relatif prima}.
			\item $TCCB=TCCB\cup HPB$.\\ See Corollary \ref{akibat-dimensi}.
		\end{itemize}
		\item
		Hence, $TCCB$ is a basis for $\mathcal{C}(A,\gamma).$
	\end{enumerate}
\end{alg}

\subsection{Examples}

Now, let us look at the three examples below.

\begin{contoh}\label{C1}

In this example, we determine a basis for and the dimension of  $\mathcal{C}(A,1).$ Let $F=\FF_3$ and  $$A=\begin{bmatrix}
		0&1&1&1\\
		1&0&1&1\\
		1&1&0&1\\
		1&1&1&0
		\end{bmatrix}.$$
		We obtain
		\begin{enumerate}
			\item $m_A(x)=x^2+x=x(x+1)$, a product of irreducible polynomials $(x)$ and $(x+1)$, and
			\item $\ker(A)=\sepan\lbrace[1\ 1\ 1\ 1]^t\rbrace$ and $\ker(A+I)=\sepan\lbrace[2\ 0\ 0\ 1]^t,[2\ 0\ 1\ 0]^t,[2\ 1\ 0\ 0]^t\rbrace$.
		\end{enumerate}
		
		Let $v_{1,1}=[1\ 1\ 1\ 1]^t,v_{2,1}=[2\ 0\ 0\ 1]^t,v_{2,2}=[2\ 0\ 1\ 0]^t,$ and $v_{2,3}=[2\ 1\ 0\ 0]^t$.
		Hence a primary cyclic decomposition of $[A]F^4$ is
		\[[A]F^4=\underbrace{\left[\langle v_{1,1}\rangle\right]}_{[A]F^4_{(x)}} \oplus\underbrace{\left[\langle v_{2,1}\rangle\oplus\langle v_{2,2}\rangle\oplus\langle v_{2,3}\rangle  \right]}_{[A]F^4_{(x+1)}}  \] where $o(v_{1,1})=x$ and $o(v_{2,1})=o(v_{2,2})=o(v_{2,3})=x+1$.
		Therefore $s(1,1)=1,s(2,1)=s(2,2)=s(2,3)=1$.
		This implies
\[
\begin{aligned}
        \dim\ndofx [A]F^4_{(x)}&=\dim\homofx(\langle v_{1,1}\rangle,\langle v_{1,1}\rangle)
		\\&=\deg (x)\min\lbrace s(1,1),s(1,1)\rbrace
		\\&=1,\ \ \ \text{and}
		\\\dim\ndofx [A]F^4_{(x+1)}&=\sum_{c=1}^3\sum_{d=1}^3\dim\homofx(\langle v_{2,c}\rangle,\langle v_{2,d}\rangle)
		\\&=\sum_{c=1}^3\sum_{d=1}^3\deg(x+1)\min\lbrace s(2,c),s(2,d)\rbrace
		\\&=9.
\end{aligned}
\]
		Therefore $\dim\mathcal{C}(A,1)=10$.
		
		Now, we will determine a basis for $\mathcal{C}(A,1)$.

The first element of the basis, say $X_1$, is the linear mapping which is the $F[x]$-endomorphism on the cyclic submodule $\langle v_{1,1} \rangle$. Hence,
\[ X_1(v_{1,1}) = v_{1,1}, \quad X_1( v_{2, j}) = 0, \ \hbox{for all} \ j =1, 2, 3.\]
The second element of the basis, say $X_2$, is the linear mapping which is the $F[x]$-endomorphism on the
cyclic submodule $\langle v_{2,1} \rangle $. The third element of the basis, $X_3$ is the linear mapping
which is the $F[x]$-homomorphism from the cyclic submodule $\langle v_{2,1} \rangle $ to $\langle v_{2,2}\rangle$.
That is
\[ X_3(v_{2,1}) = v_{2,2}, \quad     X_3(v_{1,1}) = X_3(v_{2,2}) =  X_3(v_{2,3}) = 0.\]
Continuing that process we obtain  $\{X_1, X_2, \dots, X_{10}\}$ a basis of $\mathcal{C}(A,1)$. In this case
%we obtain, for examples
%$
%X_1 	=
%		\begin{bmatrix}
%		1&1&1&1\\
%		1&1&1&1\\
%		1&1&1&1\\
%		1&1&1&1
%		\end{bmatrix},
%X_2 =	\begin{bmatrix}
%		1&1&1&0\\
%		0&0&0&0\\
%		0&0&0&0\\
%		2&2&2&0
%		\end{bmatrix},
%X_3 	=
%		\begin{bmatrix}
%		1&1&1&0\\
%		0&0&0&0\\
%		2&2&2&0\\
%		0&0&0&0
%		\end{bmatrix}.$

%\color{red}
%Here are all the elements of the basis after all.
%\begin{enumerate}
%\begin{multicols}{3}

\[
\begin{array}{ccc}

X_1=\begin{bmatrix}
1&1&1&1\\
1&1&1&1\\
1&1&1&1\\
1&1&1&1
\end{bmatrix},
&
X_2=\begin{bmatrix}
1&1&1&0\\
0&0&0&0\\
0&0&0&0\\
2&2&2&0
\end{bmatrix},
&
X_3=\begin{bmatrix}
1&1&1&0\\
0&0&0&0\\
2&2&2&0\\
0&0&0&0
\end{bmatrix},\\
\vspace{0.3cm}

X_4=\begin{bmatrix}
1&1&1&0\\
2&2&2&0\\
0&0&0&0\\
0&0&0&0
\end{bmatrix},
&
X_5=\begin{bmatrix}
1&1&0&1\\
0&0&0&0\\
0&0&0&0\\
2&2&0&2
\end{bmatrix},
&
X_6=\begin{bmatrix}
1&1&0&1\\
0&0&0&0\\
2&2&0&2\\
0&0&0&0
\end{bmatrix},\\
\vspace{0.3cm}

X_7=\begin{bmatrix}
1&1&0&1\\
2&2&0&2\\
0&0&0&0\\
0&0&0&0
\end{bmatrix},
&
X_8=\begin{bmatrix}
1&0&1&1\\
0&0&0&0\\
0&0&0&0\\
2&0&2&2
\end{bmatrix},
&
X_9=\begin{bmatrix}
1&0&1&1\\
0&0&0&0\\
2&0&2&2\\
0&0&0&0
\end{bmatrix},\\
\vspace{0.3cm}

X_{10}=\begin{bmatrix}
1&0&1&1\\
2&0&2&2\\
0&0&0&0\\
0&0&0&0
\end{bmatrix}.

\end{array}
\]
$\diamondsuit$

\end{contoh}		
		
\begin{contoh}
In this example, we determine the dimension of and a basis for $\mathcal{C}(A,2)$.
		Let $F=\FF_5$ and
		\[
		A=\begin{bmatrix}
		3&0&0&0\\
		0&3&0&0\\
		0&0&1&1\\
		0&0&0&1
		\end{bmatrix}.
	\]
		Similar to Example \ref{C1}, we obtain that a primary cyclic decomposition of $[A]F^4$ is \[[A]F^4=\underbrace{[\langle v_{1,1}\rangle\oplus\langle v_{1,2}\rangle}_{[A]F^4_{(x+2)}}]\oplus\underbrace{[\langle v_{2,1}\rangle]}_{[A]F^4_{(x+4)}} \]
		where $v_{1,1}=[1\ 0\ 0\ 0]^t,v_{1,2}=[0\ 1\ 0\ 0]^t,$ and $v_{2,1}=[0\ 0\ 0\ 1]^t$,
$o(v_{1,1})=o(v_{1,2})=x+2$ and $o(v_{2,1})=(x+4)^2$, and $s(1,1)=s(1,2)=1$ and $s(2,1)=2$. The above
decomposition is also a primary cyclic decomposition of $[2A]F^4$, with $o(v_{1,1})=o(v_{1,2})=x+4$ and $o(v_{2,1})=(x+3)^2$ and $s(1,1)=s(1,2)=1$ and $s(2,1)=2$.

		This implies
		\[
		\dim\mathcal{C}(A,2)=\dim\homofx([2A]F^4_{(x+4)},[A]F^4_{(x+4)}) = 2.
		\]
		Therefore $\dim\mathcal{C}(A,2)=2$.
		
		Now, we will determine  $\{X_1, X_2 \}$ a basis for $\mathcal{C}(A,2)$.
The first element $X_1$ is a linear mapping which is an $F[x]$-homomorphism from
 $\langle v_{1,1} \rangle$ the submodule of $[2A]F^4$ to $\langle v_{2,1}\rangle$ the
 submodule of $[A]F^4$. That is
\[
 X_1(v_{1,1}) = (A+4I)(v_{2,1}) = [0, 0, 1,0 ], \ X_1(v_{1,2}) = X_1(v_{2,1}) = X_1(2A v_{1,2})= 0.
\]
 Similarly, we obtain $X_2$ is a linear mapping
 \[
 X_1(v_{1,2}) = (A+4I)(v_{2,1}) = [0,0,1,0], \ X_1(v_{1,2}) = X_1(v_{2,1}) = X_1(2A v_{1,2})= 0.
 \]
  Hence
\[
X_1 =\begin{bmatrix}
		0&0&0&0\\
		0&0&0&0\\
		1&0&0&0\\
		0&0&0&0
		\end{bmatrix}		\text{ and }
~X_2 = \begin{bmatrix}
		0&0&0&0\\
		0&0&0&0\\
		0&1&0&0\\
		0&0&0&0
		\end{bmatrix}.
\]
$\diamondsuit$
	
\end{contoh}

\begin{contoh}
	In this example, we determine the dimension of and a basis for $\mathcal{C}(A,2)$.
	Let $F=\FF_5$ and
\[
A=\begin{bmatrix}
	1&4&2&0&4\\
	1&1&3&2&0\\
	4&4&3&4&1\\
	2&4&4&3&3\\
	3&4&1&2&1
	\end{bmatrix}.
\]
	We obtain $m_A(x)=x(x^2+2x+3)(x^2+4x+2)$ where $x,(x^2+2x+3)$ and $(x^2+4x+2)$ are irreducible in $F[x]$.	
	A primary cyclic decomposition of $[A]F^5$ is \[[A]F^5=\underbrace{[\langle v_{1,1}\rangle]}_{[A]F^5_{(x)}}\oplus\underbrace{[\langle v_{2,1}\rangle]}_{[A]F^5_{(x^2+2x+3)}}\oplus\underbrace{[\langle v_{3,1}\rangle]}_{[A]F^5_{(x^2+4x+2)}} \]
where $v_{1,1}=[1\ 1\ 4\ 3\ 3]^t,\ v_{2,1}=[1\ 0\ 1\ 1\ 1]^t,$ and $v_{3,1}=[1\ 1\ 0\ 4\ 0]^t$,
$o(v_{1,1})=x,\ o(v_{2,1})=x^2+2x+3,$ and $o(v_{3,1})=x^2+4x+2$, and $s(1,1)=1,\ s(2,1)=1,$ and $s(3,1)=1$.
	Further \[[2A]F^5=\underbrace{[\langle v_{1,1}\rangle]}_{[2A]F^5_{(x)}}\oplus\underbrace{[\langle v_{2,1}\rangle]}_{[2A]F^5_{(x^2+4x+2)}}\oplus\underbrace{[\langle v_{3,1}\rangle]}_{[2A]F^5_{(x^2+3x+3)}} \]
	where $o(v_{1,1})=x,\ o(v_{2,1})=x^2+4x+2,$ and $o(v_{3,1})=x^2+3x+3$, and $s(1,1)=1,\ s(2,1)=1,$ and $s(3,1)=1$.
	This implies
$	\dim\mathcal{C}(A,2)= 1+2=3$. A basis of
	$\mathcal{C}(A,2)$ is of the form $\{X_1,X_2, AX_2 \}$.

Now, we will determine $X_1$ and $X_2$.
 	The first matrix $X_1$ is a linear mapping which is an $F[x]$-homomorphism from
	$\langle v_{1,1} \rangle$ the submodule of $[2A]F^5$ to $\langle v_{1,1}\rangle$ the
	submodule of $[A]F^5$. That is
	\[ X_1(v_{1,1}) = v_{1,1}, \ X_1(v_{2,1})=X_1(2Av_{2,1})=X_1(v_{3,1})=X_1(2Av_{3,1})= 0.\]
	The second matrix $X_2$ is a linear mapping which is an $F[x]$-homomorphism from
	$\langle v_{2,1} \rangle$ the submodule of $[2A]F^5$ to $\langle v_{3,1}\rangle$ the
	submodule of $[A]F^5$. That is
	\[ X_2(v_{2,1}) = v_{3,1}, \ X_2(2Av_{2,1}) =Av_{3,1}=[0\ 0\ 4\ 3\ 0]^t,\  X_2(v_{1,1})=X_2(v_{3,1})=X_2(2Av_{3,1})= 0.\]

Hence
\[
X_1 = \begin{bmatrix}
	1&0&3&1&0\\
	1&0&3&1&0\\
	4&0&2&4&0\\
	3&0&4&3&0\\
	3&0&4&3&0
	\end{bmatrix},~	
	X_2=\begin{bmatrix}
	0&3&4&3&4\\
	0&3&4&3&4\\
	2&2&2&4&2\\
	4&1&0&0&0\\
	0&0&0&0&0
	\end{bmatrix},~ \text{ and }
	AX_2=\begin{bmatrix}
	4&4&4&3&4\\
	4&4&4&3&4\\
	2&4&3&1&3\\
	0&4&2&4&2\\
	0&0&0&0&0
	\end{bmatrix}.
\]
	$\diamondsuit$
	
	%\end{enumerate}
	
\end{contoh}

%Knowledge and information about basis for and dimension of twisted centralizers code give us a detail view of the code.

\section{Concluding remarks}

In this paper we have studied twisted centralizer codes and determined a basis and the dimension of the codes.
Information about the dimension of a twisted centralizer code is very crucial from the viewpoint of coding theory, since
the dimension of a linear code is one of the three important parameters of the code.
By identifying a twisted centralizer code as an $\fq[x]-$module of homomorphisms,  and also by using basic facts concerning structures of
 $\fq[x]-$module of homomorphisms we have reached our goal.
Our module theory approach has helped us in determining  a basis for and the dimension of a twisted centralizer code
algorithmically.

This paper can open a new viewpoint in dealing with generalized twisted centralizer codes defined in \cite{gtc} by $\mathcal{C}(A,D)=\lbrace X\in\fq^{n\times n}:AX=XAD\rbrace,$ where $D \in \FF_q^{n \times n}.$ As an example, it is easy to identify that a generalized twisted centralizer code $\mathcal{C}(A,D)$ is nothing but a module of homomorphisms from $\fq^n$ as $\fq[x]-$module induced by $AD$ to $\fq^n$ as $\fq[x]-$module induced by $A$. By investigating some relationship between minimal polynomials of $AD$ and $A$ we should be able to identify some structures of generalized twisted centralizer codes.
These results, which are in preparation, will be published in a separated paper.

\section*{Acknowledgements}
The authors thank two anonymous referees for their meticulous reading of the manuscript.  Their suggestions have been very valuable
 in improving the presentation of the paper.

G.P. and D.S. are supported by \emph{Institut Teknologi Bandung.}  D.S. is also supported in part by \emph{Kementerian Riset dan
Teknologi/ Badan Riset dan Inovasi Nasional (Kemenristek/ BRIN),} Republic of Indonesia.

\end{document}